# Magnetic switching by spin torque from the spin Hall effect

Luqiao Liu[1], O. J. Lee[1], T. J. Gudmundsen[1], D. C. Ralph[1,2] and R. A. Buhrman[1]

[1]Cornell University and [2]Kavli Institute at Cornell, Ithaca, New York, 14853

The spin Hall effect (SHE) generates spin currents within nonmagnetic materials. Previously, studies of the SHE have been motivated primarily to understand its fundamental origin and magnitude. Here we demonstrate, using measurement and modeling, that in a Pt/Co bilayer with perpendicular magnetic anisotropy the SHE can produce a spin transfer torque that is strong enough to efficiently rotate and reversibly switch the Co magnetization, thereby providing a new strategy both to understand the SHE and to manipulate magnets. We suggest that the SHE torque can have a similarly strong influence on current-driven magnetic domain wall motion in Pt/ferromagnet multilayers. We estimate that in optimized devices the SHE torque can switch magnetic moments using currents comparable to those in magnetic tunnel junctions operated by conventional spin-torque switching, meaning that the SHE can enable magnetic memory and logic devices with similar performance but simpler architecture than the current state of the art.

The spin Hall effect[1-6] (SHE) in Pt is strong[7-9]. Experiments show[10-12] that a charge current density $J_e$ flowing in Pt generates a transverse spin current density $(\hbar/2)J_S/e$ with $J_S/J_e > 0.05$. Some early measurements reported much smaller values[6,13,14], but for reasons that can now be understood[11,12,15]. In a thin film geometry such as shown in Fig. 1e, where the charge current flows through a small in-plane area $a$ and the spin current can act through a much larger perpendicular area $A$, the ratio of the total spin current to the total charge current $I_S/I_e = J_S A/(J_e a)$ can be much greater than one, meaning that for every electron charge passing through the device many $\hbar/2$ units of angular momentum can flow perpendicular to the film to apply a spin transfer torque to an adjacent magnetic layer. Previous experiments have shown that this SHE torque can be strong enough to excite spin wave oscillations[16], induce ferromagnetic resonance[11] or tune magnetic damping[10,11]. Here we show, by both measurement and modeling, that spin torque from the SHE of Pt can induce magnetization rotation and efficient magnetic switching in an adjacent perpendicularly-magnetized ferromagnetic layer. Our analysis shows that the SHE torque is capable of driving switching in magnetic memory devices using switching currents that are comparable to conventional spin-transfer-torque magnetic tunnel junctions[17], so that SHE-torque switching could be highly effective in technological applications.

The primary experimental phenomenon that we will analyze is room-temperature magnetic switching of a perpendicularly polarized magnetic layer within a ferromagnet/platinum bilayer driven by in-plane current. Figures 1a and 1b show an example of such switching for a Pt(20)/Co(6)/Al(16) (thicknesses in Å) multilayer patterned into a Hall bar geometry 20 μm wide and 200 μm long (Fig. 1c), with a total resistance ~ 2000 Ω. The Al layer is oxidized in air. We measure the anomalous Hall resistance, $R_H$, which is proportional to the average vertical component of the Co magnetization $M_z$[18]. Measurements as a function of vertical magnetic field near zero current establish the existence of perpendicular magnetic anisotropy (Fig. 1d). In Fig. 1a we apply a small constant in-plane magnetic field (along the current direction $\hat{y}$ as shown in Fig. 1e, *i.e.* $\beta = 0°$) that tilts the average moment by approximately 2° from vertical, but does not provide any preference for either the up or down magnetic state in the absence of current. A quasi-static in-plane current then generates hysteretic magnetic switching between the $M_z > 0$ and $M_z < 0$ magnetic states, with positive current favoring $M_z < 0$ (Fig. 1a). If the small constant in-plane magnetic field is reversed, the current-driven transitions invert, with positive current now favoring $M_z > 0$, despite the fact that the in-plane field still does not favor either magnetic state in the absence of current (Fig. 1b). The Oersted magnetic field generated by the quasistatic current cannot explain this remarkable result, as it is oriented in plane. We have observed the same switching phenomenon in Hall bars with widths varying from 1 to 20 μm, with at least five samples studied for each type. Very similar switching has also been reported recently by Miron et al.[19], who argued that the mechanism was primarily a current-generated Rashba field associated with having different materials below (Pt) and above (AlO$_x$) the Co layer. We will demonstrate, instead, that the full phase diagram as a function of current and magnetic field can be explained quantitatively by the SHE torque from the Pt layer, and that the Rashba effects proposed by Miron et al.[19,20] do not make a measurable contribution to the magnetic orientation in our samples.

*Qualitative predictions of a macrospin model.* We first solve a simple zero-temperature macrospin model to illustrate the types of behavior that can be generated by a SHE torque acting on a magnetic layer with perpendicular anisotropy. We assume the magnetic layer has thickness $t$, constant magnetization magnitude $M_S$, and magnetic orientation $\hat{m}$ and lies on top of a Pt layer of thickness $d$. We consider the geometry shown in Fig. 1e, with the sample in the $xy$ plane and with an applied magnetic field $\vec{B}_{ext} = 0\hat{x} + B_y\hat{y} + B_z\hat{z}$ (the model is generalizable for other field directions). Positive current (electrons flowing in the $-\hat{y}$ direction) induces a spin Hall current within the Pt layer such that spin moments pointing in the $\hat{\sigma} = \hat{x}$ direction (spin angular momentum pointing in $-\hat{x}$) flow upward, in the $\hat{z}$ direction. When this spin current reaches the Pt/ferromagnet interface, the spin component perpendicular to $\hat{m}$ can be absorbed by the ferromagnet, imparting a spin-transfer "torque"[21] per unit moment $\vec{\tau}_{ST} = \tau_{ST}^0 (\hat{m} \times \hat{\sigma} \times \hat{m}) = \frac{\hbar}{2eM_s t} J_S (\hat{m} \times \hat{\sigma} \times \hat{m})$. Taking into account also the torques (per unit moment) due to the external magnetic field, $\vec{\tau}_{ext} = -\hat{m} \times \vec{B}_{ext}$, and due to the anisotropy field, $\vec{\tau}_{an} = -\hat{m} \times \vec{B}_{an} = -\hat{m} \times \left[ -B_{an}^0 (\hat{m} - m_z \hat{z}) \right] = -B_{an}^0 m_z (\hat{m} \times \hat{z})$, the equilibrium orientations of $\hat{m}$ satisfy the condition $\vec{\tau}_{tot} = \vec{\tau}_{ST} + \vec{\tau}_{ext} + \vec{\tau}_{an} = 0$. We use macrospin simulations of the equation of motion[21] $(1/|\gamma|) d\hat{m}/dt = \vec{\tau}_{tot} + (\alpha/|\gamma|) \hat{m} \times (d\hat{m}/dt)$ with $\alpha > 0$ to distinguish stable from unstable equilibria.

Within this model, we calculate how the orientation of $\hat{m}$ depends on $\tau_{ST}^0$ and $\vec{B}_{ext}$. For currents corresponding to small to moderate values of spin torque, $|\tau_{ST}^0| < 0.5 B_{an}^0$, $\hat{m}$ can remain within the $yz$ plane as long as $B_x = 0$. In this case all three torques ($\vec{\tau}_{ST}, \vec{\tau}_{ext}, \vec{\tau}_{an}$) are collinear in the $\hat{x}$ direction and the torque balance equation that determines the magnetization angle $\theta$ takes a simple scalar form,

$$\tau_{tot} \equiv \hat{x} \cdot (\vec{\tau}_{ST} + \vec{\tau}_{ext} + \vec{\tau}_{an}) = \tau_{ST}^0 + B_{ext} \sin(\theta - \beta) - B_{an}^0 \sin\theta \cos\theta = 0, \qquad (1)$$

with $\theta$ and the applied field angle $\beta$ defined as in Fig. 1e with $-\pi/2 < \beta \leq \pi/2$. As the current is ramped from zero for fixed values of $B_{ext}$ and $\beta$, initially the dominant effect of $\vec{\tau}_{ST}$ is not to provide an anti-damping torque as it would if the Co were magnetized in-plane, but rather to rotate $\hat{m}$ within the *yz* plane, thereby shifting $\theta$ continuously, until, for sufficiently large currents, Eq. (1) predicts abrupt hysteretic switching. In Figure 2a, we show magnetic hysteresis curves predicted by this macrospin model for representative fixed values of the in-plane magnetic field. Just as observed experimentally in Figs. 1a,b, the sign of the hysteresis reverses when the in-plane field component is reversed. The reason for this reversal is that although an in-plane magnetic field does not favor either magnetic orientation by itself, an in-plane field breaks the symmetry in the response to the SHE torque. With a magnetic field component in the in-plane $\hat{y}$ direction, the barrier against clockwise rotation of $\hat{m}$ from the $m_z > 0$ state to the $m_z < 0$ is different than for clockwise rotation from the $m_z < 0$ to the $m_z > 0$ state, with the result that the direction of the in-plane field determines which out-of-plane magnetic orientation will be favored by a given sign of the SHE torque (Fig. 2b).

For very large spin torques, $|\tau_{ST}^0| > B_{an}^0 / 2$, the SHE torque is greater than the maximum value of the restoring torque from the magnetic anisotropy, $|\vec{\tau}_{an}|$, and for sufficiently small values of $|B_{ext}|$ there is no solution of Eq. (1). This means that $\hat{m}$ cannot remain in the *yz* plane. By solving the full vector equation $\vec{\tau}_{tot} = 0$, we find that for large $|\tau_{ST}^0|$ there is a current-stabilized state in which $\hat{m}$ develops a component in the $+\hat{x}$ direction for positive $\tau_{ST}^0$ and $\hat{m}$ tilts toward $-\hat{x}$ for negative $\tau_{ST}^0$.

The full phase diagram in the macrospin model for $\hat{m}(\tau_{ST}^0, B_y, B_z)$ can be calculated as described in the Supplementary Information. We show particular sections through the phase diagram in Fig. 3a,b. In the central areas of both Fig. 3a and 3b (labeled $m_z = \uparrow / \downarrow$, $m_x = 0$) $\hat{m}$ is bistable between the $\{m_z > 0, m_x = 0\}$ and $\{m_z < 0, m_x = 0\}$ states. The solid lines are boundaries at which one of these $m_x = 0$ states is destabilized, producing a hysteretic transition. These switching boundaries can be calculated analytically by incorporating $\tau_{ST}^0$ into standard Stoner-Wohlfarth theory[22], using Eq. (1) together with the condition $d\tau_{tot}/d\theta = 0$, with a result that can be expressed in a symmetric form describing switching conditions for effective magnetic field components in the y and z directions

$$B_y + \tau_{ST}^0 \sin\theta = B_{an}^0 \cos^3\theta \qquad (2)$$

$$B_z - \tau_{ST}^0 \cos\theta = -B_{an}^0 \sin^3\theta. \qquad (3)$$

Equations (2) and (3) allow a numerical solution for the value of the SHE torque needed to achieve switching for any fixed values of $B_y$ and $B_z$. The distinct high current states for which the SHE torque tilts the magnetic orientation out of the yz plane are labeled in Figs. 3a,b by $m_x > 0$ and $m_x < 0$. When $\left|\tau_{ST}^0\right| > B_{an}^0/2$ these $m_x \neq 0$ states are the only allowed solutions for small $\left|B_{ext}\right|$. Once formed, the $m_x \neq 0$ states can remain stable even for a range of smaller $\left|\tau_{ST}^0\right|$. The dashed lines in Figs. 3a,b represent the boundaries at which the $m_x \neq 0$ states become locally unstable.

*Measurements of the SHE torque and the Rashba field.* Before analyzing the switching data, we consider the regime in which the Co magnetic moment rotates coherently. By comparing the direction and magnitude of rotation induced by current to those of rotations induced by magnetic fields applied in different directions we can distinguish the SHE torque from a conventional Rashba field[23], and measure the SHE torque.

We first apply $\vec{B}_{ext}$ in the yz plane with a small angle $\beta = 4°$ relative to the y axis (Fig. 2c). In this case the field-induced torque is parallel to $\hat{x}$ so it adds to or subtracts from a SHE torque, depending on the sign of *I*. The nonzero angle $\beta$ suppresses domain formation so that the magnetization rotates coherently, and the macrospin model applies. We compare field sweeps for the same magnitude of current, positive and negative (±12 mA in Fig. 2c), so that the consequences of Ohmic heating should be identical for both. We define $B_+(\theta)$ as the value of $B_{ext}$ required to produce a given value of $\theta$ when $I$ is positive and $B_-(\theta)$ as the corresponding quantity for $I$ negative. From Eq. (1), $B_{+/-}(\theta) = [B_{an}^0 \sin\theta \cos\theta \mp \tau_{ST}^0]/\sin(\theta - \beta)$, so that $B_-(\theta) - B_+(\theta) = 2\tau_{ST}^0/\sin(\theta - \beta)$. The angle $\beta$ is known for our apparatus with an accuracy of ±1° (see Methods) and $\sin\theta$ can be determined accurately from $R_H$. Therefore, by taking the difference of the two experimental $B_{ext}$ vs. $R_H$ curves (for ± $I$) (Fig. 2d) and performing a one-parameter fit, we can determine $\tau_{ST}^0 = 4.0 \pm 0.7$ mT for $I = 12$ mA, or $\tau_{ST}^0/I = 0.33 \pm 0.06$ mT/mA. We find that $\tau_{ST}^0/I$ is approximately independent of $I$ (Fig. 2e). A current of 12 mA corresponds to a charge current density $J_e = 2.3 \times 10^7$ A/cm$^2$, assuming for simplicity that the current density is uniform throughout the Pt/Co bilayer and the Al is fully oxidized. Using $\tau_{ST}^0 = \hbar J_S/(2eM_S t)$ with $M_S \approx 1.0 \times 10^6$ A/m measured for our Co films, our value of $\tau_{ST}^0$ at 12 mA corresponds to $J_S \approx 7 \times 10^5$ A/cm$^2$, or $J_S(d = 2 \text{ nm})/J_e = 0.03 \pm 0.01$. After accounting for a correction associated with the fact that the thickness $d$ of our Pt layer is comparable to the spin diffusion length in Pt, this value of $J_S(d = 2 \text{ nm})/J_e$ corresponds to a bulk value $J_S(d = \infty)/J_e = 0.06 \pm 0.02$ (see Supplementary Information), in quantitative agreement with previous measurements on Pt/permalloy bilayers[10-12,15]. A similar analysis of $B_+(\theta) + B_-(\theta)$ allows a determination of $B_{an}^0$ as a function of $|I|$ (see Supplementary Information). We find $B_{an}^0 = 280$ mT near $I = 0$ and decreases significantly as a function of increasing $|I|$, reflecting strong heating.

Next we describe a similar experiment with $\vec{B}_{ext} = B_x \hat{x}$. If there is a current-induced Rashba field, it should be primarily in the $\hat{x}$ direction and therefore it should cause current-induced shifts in $R_H$ vs. $B_x$ curves. Figure 2f shows representative data for $I = \pm 10$ mA, corresponding to a current density $1.9 \times 10^7$ A/cm² in the 20 μm wide sample. We observe no measurable shift between the two curves for any value of $|I|$, from which we conclude that any Rashba field in our sample has a magnitude that is less than our sensitivity, $|B_{Rashba}|/J_e < 1.3 \times 10^{-7}$ mT /(A/cm²). This result is in striking contrast to a report by Miron et al., that a Rashba field (parallel to $\hat{x}$) 75 times larger than our upper bound exists for similar Pt(30)/Co(6)/AlO$_x$ samples[20]. Our null result is consistent with the Oersted field in our samples, which is in the $\pm\hat{x}$ direction but which has a magnitude less than our sensitivity; $|B_{Oersted}|/J_e = \mu_0 d/2 = 1.3 \times 10^{-8}$ mT /(A/cm²) by Ampere's law.

*Analysis of the experimental switching phase diagrams.* We have assembled experimental switching phase diagrams (SPDs) by measuring switching transitions as a function of $I$, $B_y$, and $B_z$. Representative sections through the SPDs are plotted in Figs. 3c,d. Qualitatively, we note that these SPDs have shapes and symmetries very similar to the boundaries of the $m_z = \uparrow/\downarrow$, $m_x = 0$ states in the macrospin model (Figures 3a,b), supporting our assertion that the switching can be explained by the SHE torque. However, to analyze the effects of the SHE torque quantitatively, it is not appropriate to use just a zero-temperature macrospin model for two reasons: (i) current-induced heating can be significant and (ii) magnetic switching occurs by means of a spatially non-uniform reversal process. Nonuniform switching is evident even for $I = 0$, in that the switching field for perpendicularly applied fields ( $B_c = 17$ mT, see Fig. 1d) is much less than the value $B_c = B_{an}^0$ expected within the macrospin model ( $B_{an}^0 = 280$ mT near $I = 0$, determined above). Nevertheless, we can achieve a reasonable quantitative modeling of the SPDs by including the effects of the SHE torque within a modified Stoner-Wohlfarth model[24], that substitutes a reduced coercive field $B_c(|I|)$ in place of $B_{an}^0$ in Eq. (3) [but not Eq. (2)] to account approximately for the reduced switching threshold for effective fields in the $z$ direction. We determine $B_c(|I|)$ experimentally by measuring the switching field as a function of $I$ for $\vec{B}_{ext}$ perpendicular to the sample plane, the field angle for which spin torque effects are weakest (see Supplementary Information). The only other parameters in the model are the SHE torque strength $\tau_{ST}^0(I) = (0.33 \text{ mT/mA})I$ determined above and $B_{an}^0(|I|)$ determined independently by fitting to $B_+(\theta) + B_-(\theta)$ in the coherent rotation regime. With these inputs, switching currents can be calculated in the modified Stoner-Wohlfarth model for all field values and compared to the experiment with no adjustment of fitting parameters (solid lines in Figs. 3c, d). We find remarkable agreement considering the simplicity of the model. In particular, the skewed shape of the hysteretic region in Fig. 3d is reproduced with no fitting parameters. We conclude that the SHE torque in combination with heating provides a *quantitative* description for the current-driven switching we measure. Heating alone cannot explain the data, since heating depends on $|I|$ and we measure opposite signs of switching for opposite signs of $I$.

As noted above, the current-induced magnetization switching of perpendicularly magnetized layers that we describe is similar to a recent discovery by Miron *et al.*[19] who, while considering the influence of a SHE torque, argued that the mechanism of switching is due to a Rashba-induced field along $\pm\hat{z}$ (distinct from the conventional Rashba direction $\pm\hat{x}$). Because our measurements of both coherent magnetization rotation and magnetization switching can be explained *quantitatively* by the same value of the SHE torque, and this SHE torque also corresponds to a value of spin current $J_s/J_e$ in agreement with previous experiments, we argue that the SHE torque mechanism fully explains the current-induced switching in our devices, with no evidence of the new Rashba effect suggested in ref. 19 (see additional discussion in the Supplementary Information). The lack of a measurable Rashba field in the $\pm\hat{x}$ direction in our coherent rotation experiments gives additional reason to question the existence of a large Rashba field in the unconventional $\pm\hat{z}$ direction. We have also measured similar current-induced switching in Pt(30)/Co(5)/Ni(10)/Ta(10) (Fig. 4), Pt(30)/Co(5)/Ni(10)/Au(10) and Pt(30)/CoFeB(10)/MgO(16) samples (thicknesses in Å). This shows that the switching does not depend on the presence of an oxide capping layer, and occurs for ferromagnet thicknesses up to 15 Å and for ferromagnets with different chemical compositions. These observations suggest strongly that it is the Pt film which drives switching, rather than a Rashba field in the magnet whose existence and magnitude would be sensitive to these factors.

We are not able to apply large enough steady-state currents to identify unambiguously the out-of-plane $m_x \neq 0$ state predicted by the macrospin model for the high-current regime $|\tau_{ST}^0| > B_{an}^0/2$. However, this regime may be achievable using pulsed-current measurements as in ref. 20. In fact, the rotation of $\hat{m}$ out of the $yz$ plane predicted by the macrospin model in this regime has the correct symmetry to explain the observations of stochastic domain reversal reported in ref. 20, as an alternative to the Rashba analysis that led Miron et al. to claim the existence of large $\pm\hat{x}$ Rashba fields.

*Ramifications*. The realization that spin torque from the SHE is strong in Pt/ferromagnet bilayers also has important consequences for studies of current-driven magnetic domain wall motion in nanowires made from layered Pt/ferromagnet structures[25-31]. The effect of the SHE torque on a domain wall may help to explain, *e.g.*, why the non-adiabatic torque measured in Pt/ferromagnet nanowires is anomalously strong[26,28,29].

As already noted by Miron et al.[19], the discovery that an in-plane current can switch the magnetic moment of a perpendicularly-polarized magnetic film is particularly exciting because it opens a new strategy for controlling magnetic memories and non-volatile logic elements. Understanding that the SHE torque is the mechanism of switching allows us to make quantitative estimates for how to optimize the effect. For a sufficiently small sample, the macrospin model should apply. We assume a magnetic layer of length $L$ (in the current direction), width $w$, and thickness $t$ for which the perpendicular anisotropy field is optimized to provide an energy barrier of 40 $k_B T$ (where $k_B$ is Boltzmann's constant and $T$ = 300 K), corresponding to a retention time of 10 years[32]. The small, fixed, symmetry-breaking in-plane magnetic field needed to set the direction of the spin-Hall switching can be applied easily by the dipole field from a nearby magnetic layer. A simple analysis yields a critical current for SHE switching (see Supplementary Information)

$$I_c = \frac{2e(40\,k_B T)\left[d + (\sigma_F/\sigma_{Pt})t\right]}{\hbar L \left(J_S(d=\infty)/J_{e,Pt}\right)\left[1-\text{sech}(d/\lambda_{sf})\right]} \frac{M_S(|I_c|) B_{an}^0(|I_c|)}{M_S(I=0) B_{an}^0(I=0)}. \quad (4)$$

Here $d$ is the Pt thickness, $\sigma_F$ is the electrical conductivity of the ferromagnet, $\sigma_{Pt}$ is the conductivity of the Pt, and $\lambda_{sf}$ is the Pt spin diffusion length. For a sample with $L$ = 200 nm, $d$ = 2 nm, $t$ = 0.6 nm $J_S(d=\infty)/J_{e,Pt} = 0.07$[11], $\lambda_{sf}$ = 1.4 nm[15] and assuming for simplicity $\sigma_F = \sigma_{Pt}$, we conclude that $I_c$ should be ~ 170 µA even in the absence of any assistance from heating-induced thermal activation. The critical currents would be reduced even further with heating, or by using materials that might generate stronger SHE torques than our Pt. Switching currents for the SHE torque therefore have the potential to be competitive with the optimum switching currents for magnetic tunnel junctions (MTJs) controlled by conventional spin transfer torque[32-34]. Compared to conventional MTJ memory elements, spin-Hall switched devices have an advantageous architecture because charge currents flow only laterally within the device and do not need to flow through tunnel barriers that are sensitive to electrical breakdown.

**Methods**

Multilayer films with structures, from bottom to top, Pt(20)/Co(6)/Al(16), Pt(30)/Co(5)/Ni(10)/Ta(10), Pt(30)/Co(5)/Ni(10)/Au(10) and Pt (30)/CoFeB(10)/MgO(16) (thicknesses in Å) were deposited by sputtering onto thermally oxidized Si wafers at a base pressure lower than $2 \times 10^{-8}$ Torr. The growth rate was controlled to be less than 0.5 Å/s in order to achieve a highly oriented texture. The Al capping layer was oxidized by exposure to the atmosphere; no plasma oxidation was employed. The Pt/Co/Al and Pt/CoFeB/MgO samples were annealed under ultra-high vacuum at 350 °C for 1 hr, which improved the squareness of magnetic hysteresis loops as a function of magnetic field swept in the $z$ direction. No annealing was applied to the other two types of films. The films were patterned into Hall bar geometries using photolithography and ion milling. Ti/Au electrodes were evaporated to provide electrical connection.

For our Hall resistance measurements, the sample chip was installed onto a rotary stage which provided a 360° rotation range and 0.02° rotation precision. The zero point for the angle $\beta$ was determined by rotating the stage until the magnetization curve for small current flipped its polarity (see Supplementary Information). The accuracy of the angle read from the stage is limited by the uncertainty of the zero point, and is within ±1°.

**Acknowledgements**
We acknowledge support from ARO, ONR, DARPA, NSF/MRSEC (DMR-1120296) through the Cornell Center for Materials Research (CCMR), and NSF/NSEC through the Cornell Center for Nanoscale Systems. We also acknowledge NSF support through a fellowship for T.G. and through use of the Cornell Nanofabrication Facility/NNIN and the CCMR facilities.

**Author Contributions**
L. L. designed and carried out the experiment. L. L., D. R. and R. B. analyzed the data. R. B. and D. R. supervised the experiment. L. L. and D. R. drafted the manuscript. All authors contributed to discussions during the modeling and interpretation of the experiments, and all participated in the editing and finalization of the manuscript.

**Additional information**
The authors declare no competing financial interests. Correspondence and requests for materials should be addressed to Luqiao Liu.

**FIGURE CAPTIONS**

**Figure 1. Current-induced switching of a magnetic layer with perpendicular anisotropy driven by spin torque from the spin Hall effect. a,b**, Current-induced switching in a Pt/Co/AlO$_x$ sample at room temperature detected by measuring the Hall resistance $R_H$ in the presence of a small, fixed in-plane magnetic field $B_y$ with (**a**) $B_y$ = 10 mT and (**b**) $B_y$ = -10 mT. The sign of the in-plane field $B_y$ determines the direction of the current-induced switching. **c**, Top-view image of the Pt/Co/AlO$_x$ Hall bar connected with Ti/Au electrodes. The scale bar is 50 µm. **d**, $R_H$ as a function of $B_{ext}$ when $B_{ext}$ is applied perpendicular to the sample plane, along the magnetic easy axis. **e**, Schematic illustration of the fields and torques exerted on the magnetization $M$ for the case $\left|\tau_{ST}^0\right| < 0.5 B_{an}^0$ when $M$ lies in the $yz$ plane. The external field $\vec{B}_{ext}$ is applied in the $yz$ plane and forms an angle $\beta$ with the $y$ axis. $\vec{\tau}_{ext}$ and $\vec{\tau}_{an}$ represent the torques generated by $\vec{B}_{ext}$ and the anisotropy field $\vec{B}_{an}$. $\vec{\tau}_{ST}$ is the spin torque due to the SHE when electrons flow in the -$y$ direction.

**Figure 2. Predictions of a macrospin model for SHE-torque-driven magnetic dynamics, and coherent rotation experiments used to measure the magnitude of the spin Hall torque and the Rashba field. a**, Predictions for current-induced magnetic switching within the macrospin model. The red, green and blue curves correspond to in-plane magnetic fields $B_y = \pm 0.1\, B_{an}^0$, $\pm 0.2\, B_{an}^0$ and $\pm 0.4\, B_{an}^0$, respectively, and the direction of switching reverses when $B_y$ changes sign. **b**, Schematic illustration of the magnetization vectors for the two tilted magnetic states which are stable in the absence of current when a fixed in-plane magnetic field $B_y > 0$ (left) or $B_y < 0$ (right) is applied. (The angle of the tilt from vertical is exaggerated compared to the data in Figs. 1a and 1b.) The directions of current-induced switching depend on the sign of $\tau_{ST}^0$ as shown. **c**, $R_H$ vs.

$B_{ext}$ determined experimentally for a Pt/Co/AlO$_x$ sample when the magnetic field is in the $yz$ plane applied at the angle $\beta = 4°$. Constant currents of ±12 mA were applied to the sample while sweeping the field. **d**, Points: Measured values of $B_-(\theta) - B_+(\theta)$ and $[B_-(\theta) + B_+(\theta)]/2$ as defined in the text, determined from the magnetization curves of (**c**). Lines: fits to the macrospin model to determine $\tau_{ST}^0(I)$ and $B_{an}^0(I)$. **e**, The values of $\tau_{ST}^0/I$ measured for different values of $I$. The error bars are dominated by the uncertainty in the angle $\beta$. **f**, $R_H$ as a function of applied field when $B_{ext}$ is applied along the $x$ direction, measured for $I = \pm 10$ mA. The curves are indistinguishable, allowing us to set a limit on the strength of the Rashba field in our samples.

**Figure 3. Phase diagrams for the magnetic moment in the presence of the SHE torque. a**, **b**, Phase diagram calculated in the zero-temperature macrospin model for (**a**) $B_{ext}$ applied along the $y$ axis and (**b**) $B_y$ fixed at $0.2\, B_{an}^0$ with $B_z$ varied continuously. The solid lines represent the switching boundaries for the {$m_z = \uparrow/\downarrow$, $m_x = 0$} states and the dashed lines represent limits of stability for the $m_x \neq 0$ states. **c** and **d**, Phase diagram of a Pt/Co/AlO$_x$ sample determined experimentally by (**c**) sweeping $I$ for fixed values of $B_{ext}$ applied along the $y$ axis, and (**d**) fixing $B_y = 40$ mT and sweeping $B_z$ for fixed values of $I$. The solid lines in (**c**) and (**d**) represent theoretical switching boundaries calculated using the modified Stoner-Wohlfarth model. In all panels, the symbol $\uparrow$ means $m_z > 0$ and $\downarrow$ means $m_z < 0$, not $m_z = \pm 1$.

**Figure 4. Current-induced switching in a Pt/Co/Ni/Ta sample.** SHE switching in a Pt/Co/Ni/Ta device 3 μm wide and 3 μm long, with an in-plane magnetic field (**a**) $B_y = 35$ mT and (**b**) $B_y = -35$ mT. The magnitude of $R_H$ is smaller than in Fig. 1 because the scale of the anomalous Hall effect differs depending on the composition of the magnetic layer. The current-induced switching is gradual, rather than abrupt, consistent with the easy axis magnetization curves of this sample for which the field-driven switching is gradual, as well.

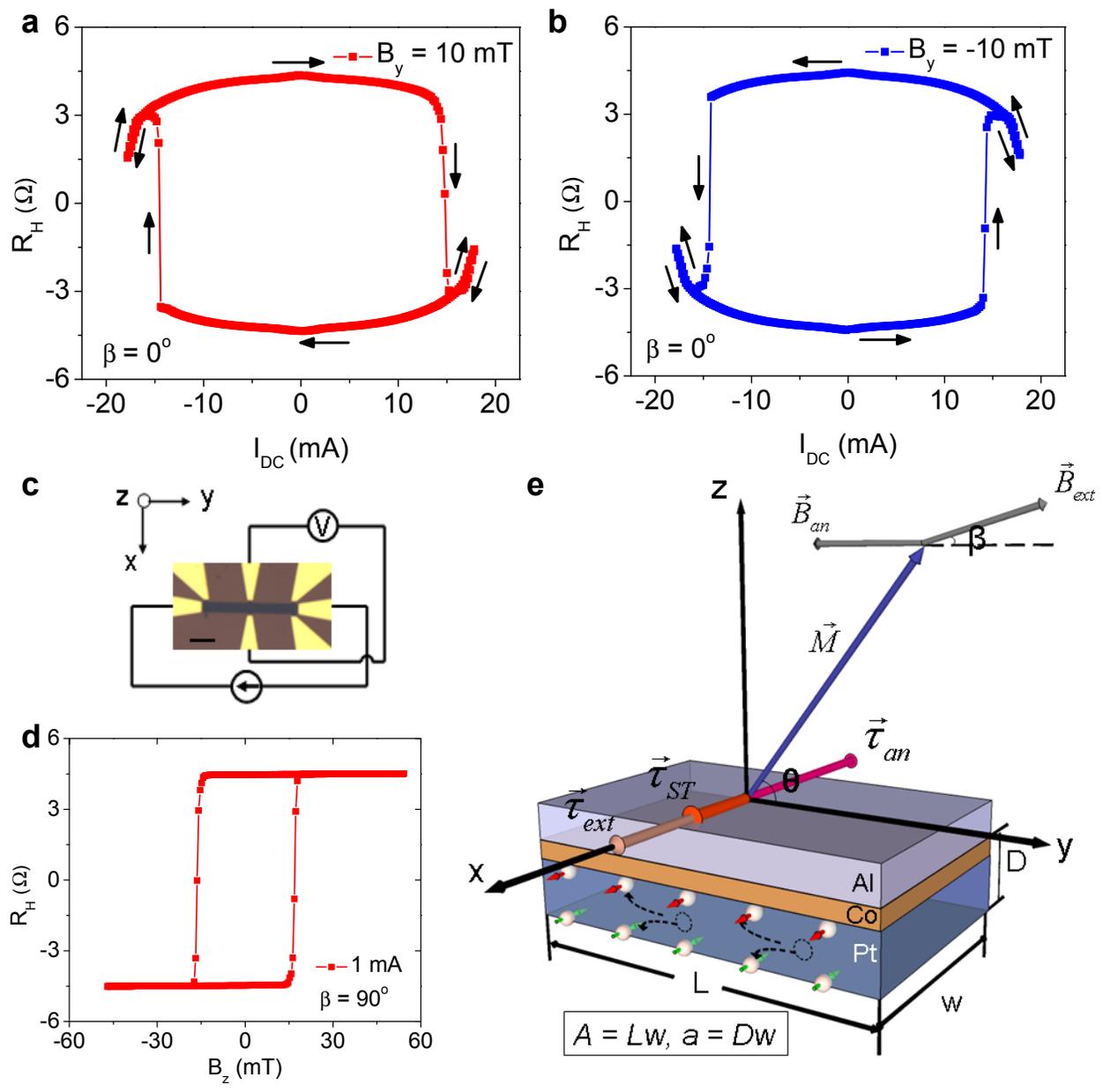

**Figure 1**

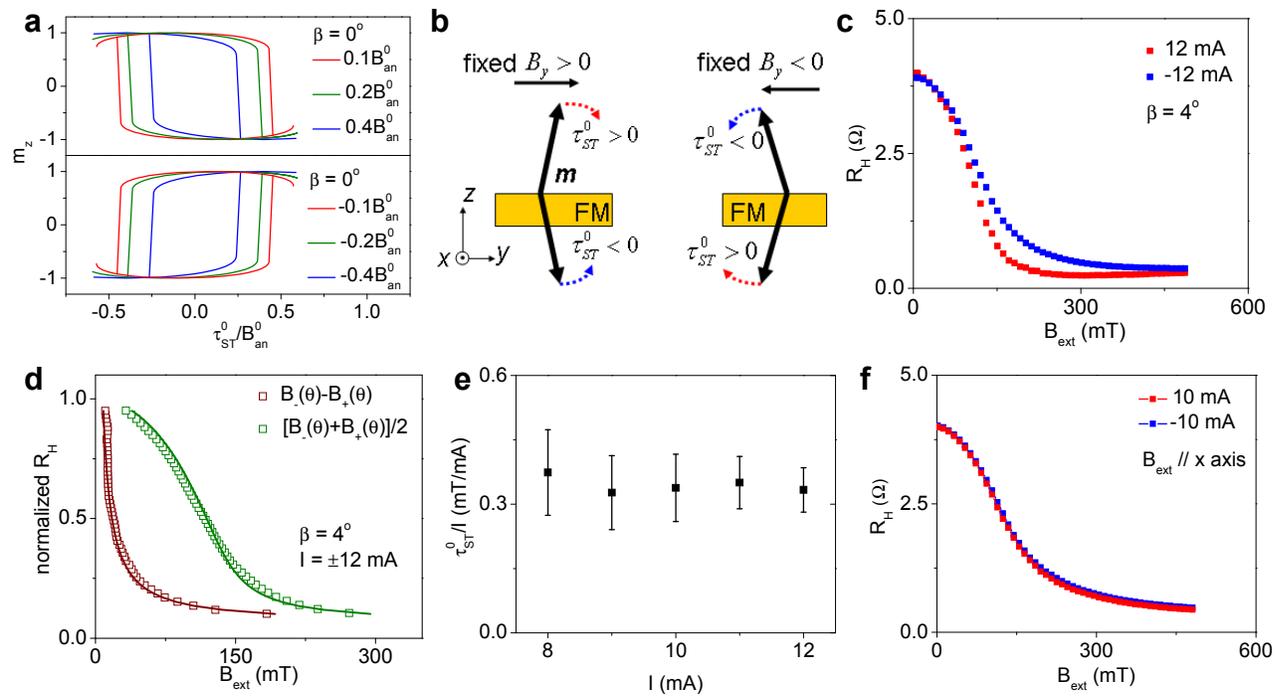

**Figure 2**

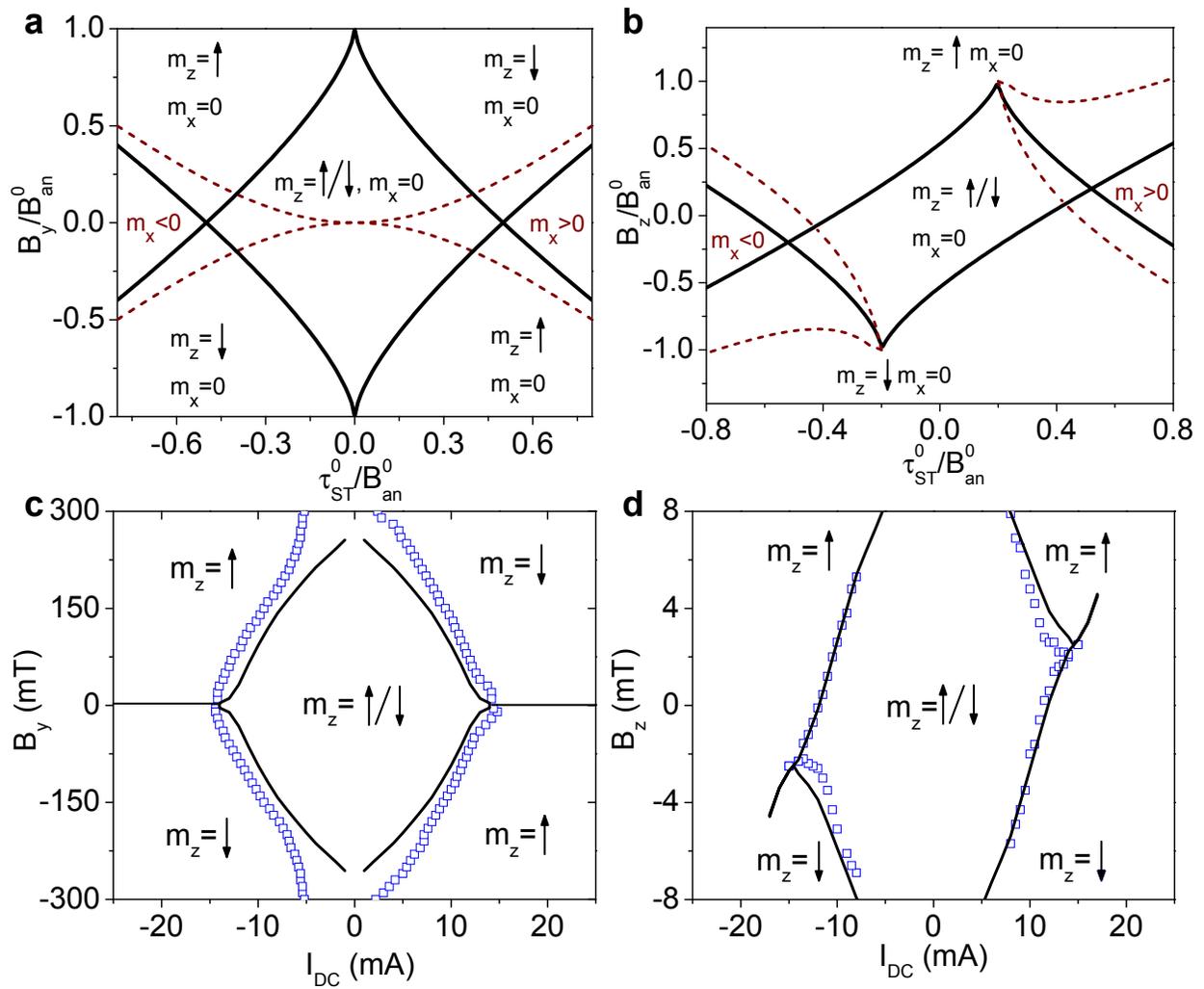

**Figure 3**

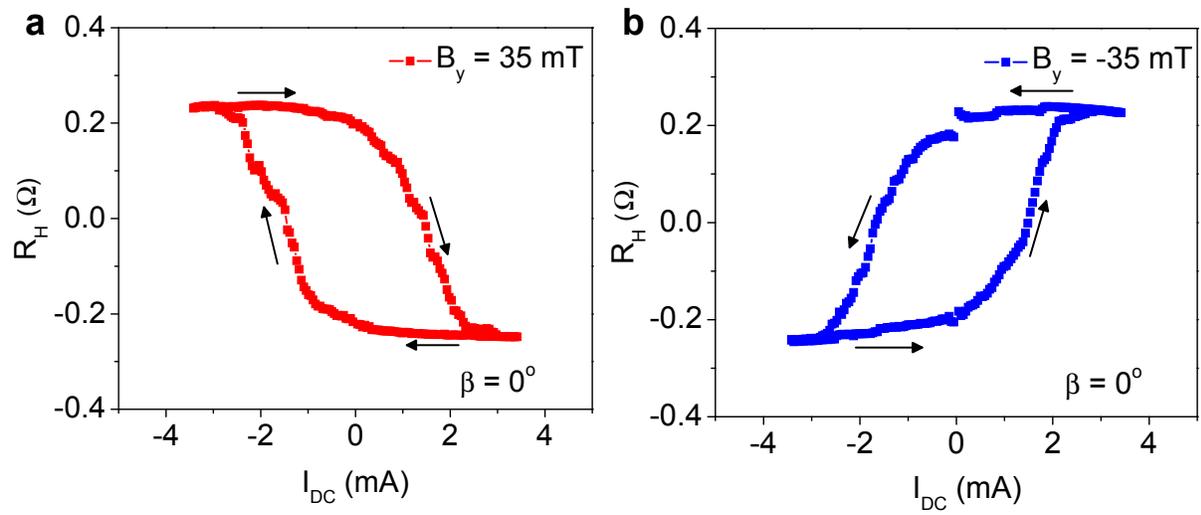

**Figure 4**

# Magnetic switching by spin torque from the spin Hall effect

Luqiao Liu[1], O. J. Lee[1], T. J. Gudmundsen[1], D. C. Ralph[1,2] and R. A. Buhrman[1]

[1]Cornell University and [2]Kavli Institute at Cornell, Ithaca, New York, 14853

## SUPPLEMENTARY INFORMATION

**Contents:**

S1. Comparing our measured value of $J_S/J_e$ to previous experiments: accounting for the Pt layer thickness.

S2. Current dependence of the perpendicular anisotropy field $B_{an}^0(|I|)$ and the coercive field $B_c(|I|)$.

S3. Estimates for the critical current for spin Hall switching of a magnetic memory element.

S4. Comments on arguments by Miron et al. in ref. S1

S5. Determining the zero point for the applied magnetic field angle $\beta$.

S6. Determining the phase diagram in the macrospin model for a spin Hall torque acting on a magnetic layer with perpendicular magnetic anisotropy.

## S1. Comparing our measured value of $J_S/J_e$ to previous experiments: accounting for the Pt layer thickness.

The measurement discussed in the main text yielded the value for our samples $J_S(d = 2 \text{ nm})/J_e = 0.03 \pm 0.01$, where $J_S$ is the spin current density (multiplied by $2e/\hbar$) penetrating from the Pt to the Co layer, and $J_e$ is the charge current density within the Pt layer. Within a bulk sample $J_S/J_e$ is known as the spin Hall angle. In order to compare our value to previous measurements[S2-S5], one should account for a correction due to the small thickness of our Pt layer ($d$ = 2.0 nm). If one assumes that no spin current can penetrate out through the bottom surface of the Pt layer, within a drift-diffusion theory for spin flow[S6] that incorporates the spin Hall effect there must arise a vertical gradient in the spin-dependent electron chemical potentials adjacent to the bottom surface (within a length scale given by the spin diffusion length $\lambda_{sf}$) that produces a counterflowing spin current to cancel the spin-Hall-generated spin current at the bottom surface. If we assume also that the component of the spin Hall current with spin oriented perpendicular to the Co magnetization is completely absorbed at the Pt/Co interface[S7] (so that there is no corresponding generation of spin-dependent chemical potential gradients at this interface affecting the component of the spin current that exerts a torque on the Co), then the prediction of the drift-diffusion theory is that the spin Hall current density measured by our technique using a Pt film of thickness $d$ should be reduced from the bulk value according to the formula $J_S(d)/J_S(\infty) = 1 - \text{sech}(d/\lambda_{sf})$ (ref. S3). Using measurements of spin-Hall-torque-driven magnetic resonance[S3] in Pt/Permalloy samples as a function of Pt thickness, we have determined that $\lambda_{sf}$ = 1.4 ± 0.3 nm at room temperature for our Pt films[S5]. Using this value, our measurement of $J_S(d = 2 \text{ nm})/J_e = 0.03 \pm 0.01$ corresponds to a corrected bulk value $J_S(d=\infty)/J_e = 0.06 \pm 0.02$, consistent with the extrapolated bulk values $J_S(d=\infty)/J_e = 0.068 \pm 0.005$ that we determined previously for our Pt films from from measurements on Pt/Permalloy samples[S3,S5].

Since no plasma oxidation is involved in the sample preparation, there is also the possibility that the Al capping layer is only partially oxidized and shunts some of the current. If this is the case, $J_e$ flowing inside the Pt layer will be smaller compared to our calculation above, and this will lead to an even larger $J_S/J_e$ value. We can determine an upper limit for $J_S/J_e$ by assuming a fully metallic Al layer. In this case $J_S(d=\infty)/J_e = 0.09 \pm 0.02$, which is still consistent with previous Pt/Permalloy results within the accuracy of the experiments.

## S2. Current dependence of the perpendicular anisotropy field $B_{an}^0(|I|)$ and the coercive field $B_c(|I|)$.

We measure the perpendicular magnetic anisotropy field $B_{an}^0(|I|)$ in our samples by considering the quantity $B_+(\theta) + B_-(\theta)$ in field ranges where the sample magnetization is saturated and rotates coherently. Here $B_+(\theta)$ is the value of $B_{ext}$ required to produce a given value of magnetization angle $\theta$ for a given positive current and field angle $\beta$, and $B_-(\theta)$ is the corresponding quantity for the corresponding value of negative current at the same values of $\theta$ and $\beta$. Based on Eq. (1) of the main text, we expect that $B_+(\theta) + B_-(\theta) = 2B_{an}^0 \sin\theta\cos\theta / \sin(\theta-\beta)$. By performing a one-parameter fit using this equation to data such as in Fig. 2d of the main text, for different values of current $|I|$, we determine $B_{an}^0(|I|)$ as plotted in Fig. S1a. We observe a strong decrease in $B_{an}^0$ as a function of $|I|$, presumably due to Ohmic heating.

In order to determine the current dependence of the coercive field, $B_c(|I|)$, we first measure the current dependence of the switching field, $B_z^{sw}$, for the external applied field oriented perpendicular to the film plane, $\beta = 90°$ (Fig. S1b). This is the field direction for which switching is least affected by the spin Hall torque. We then correct for the small influence of the spin Hall torque for this orientation within the modified Stoner-Wohlfarth model introduced in the main text:

$$B_z^{sw} - \tau_{ST}^0 \cos\theta = -B_c \sin^3\theta \tag{S1}$$

$$B_y^{sw} + \tau_{ST}^0 \sin\theta = B_{an}^0 \cos^3\theta. \tag{S2}$$

Here $B_{an}^0(|I|)$ is the anisotropy field determined as described above and $B_c(|I|)$ is the intrinsic coercive field we wish to determine. Note that for the case $\beta = 90°$ (i.e., $B_y^{sw} = 0$), as long as $\tau_{ST}^0 \ll B_{an}^0$ then from Eq. (S2) the magnetization angle $\theta$ will be near $\pm\pi/2$ at the point of switching. Therefore the term $\tau_{ST}^0 \cos(\theta)$ in Eq. (S1) will have a small magnitude, and consequently the effect of $\tau_{ST}^0$ on the switching field $B_z^{sw}$ will also be small (but not zero).

Using $B_{an}^0(|I|)$ as measured above and the value of $\tau_{ST}^0$ determined in the main text, $\tau_{ST}^0 / I = 0.33 \pm 0.06$ mT/mA, we determine $B_c(|I|)$ by performing a one-parameter fit of the numerical solution of Equations (S1) and (S2) to the measured switching fields $B_z^{sw}(I)$. The results are shown as the solid curve in Fig. S1b. As expected for $\beta = 90°$, the intrinsic coercive field $B_c(|I|)$ is only slightly larger than the measured switching fields $B_z^{sw}(|I|)$. We can see from Fig. S1b that $B_c(|I|)$, like $B_{an}^0(|I|)$, has a strong current dependence that we likewise attribute to heating.

We assume that the intrinsic coercive field $B_c(|I|)$ can be taken to be independent of the direction of the applied magnetic field within the context of the modified Stoner-Wohlfarth model. Therefore we use the same values of $B_{an}^0(|I|)$ and $B_c(|I|)$ when applying the model for other magnetic field orientations (other than $\beta = 90°$), for which the influence of the spin Hall torque is stronger. This allows us to compare our measurements to the predicted consequences of the spin Hall torque without any additional adjustment of parameters.

**S3. Estimates for the critical current for spin Hall switching of a magnetic memory element.**

We wish to estimate the minimum critical current for spin Hall switching in an optimized memory device subject to the constraint that the memory is thermally stable, i.e., that the memory has an energy activation barrier $\geq 40\ k_B T$ in the absence of a current, where $k_B$ is Boltzmann's constant and $T$ is room temperature, corresponding to a retention time of at least 10 years[S8]. As in the main text, we assume that the magnetic layer has a perpendicular anisotropy, current flows in the $y$ direction, a magnetic field $B_y$ may be applied in $\hat{y}$ direction, and the rotation angle of the magnetization relative to the sample plane is $\theta$ (see Fig. 1 of the main text). For simplicity, our estimate will use the standard (unmodified) Stoner-Wohlfarth model that assumes macrospin dynamics even during the switching process and takes into account the effects of the spin Hall torque using Equations (2) and (3) of the main text. Samples for which switching occurs by domain wall nucleation and motion will exhibit reductions in both the critical current for switching and the energy barrier, such that their ratio might be larger or smaller than our estimate.

Within this macrospin model, both the critical current for switching and the energy barrier for reversal depend on $B_y$. We therefore first consider how to set $B_y$ to minimize the critical current for a given fixed value of the energy barrier. For a thin film magnet with uniaxial anisotropy, the free energy has the form $f(B_y, \theta) = M_S A t \left[ (B_{an}^0 \cos^2 \theta)/2 - B_y \cos\theta \right]$ and the equilibrium orientation is $\theta_0 = \cos^{-1}(B_y / B_{an}^0)$. Here $M_S$ is the saturation magnetization, $A$ is sample area, and $t$ the thickness of the ferromagnetic layer. Therefore the energy barrier is

$$\Delta f(B_y) = f(B_y, \theta = \theta_0) - f(B_y, \theta = 0) = M_S A t B_{an}^0 [1 - (B_y / B_{an}^0)]^2 / 2. \tag{S3}$$

In order to determine the switching current as a function of $B_y$ for a given energy barrier, for each value of $B_y$ we scale the parameter $B_{an}^0$ in Eqs. (2) and (3) of the main text so that the energy barrier is constant, and then use Eqs. (2) and (3) of the main text to calculate the switching current. Our result for the switching current as a function of $B_y$ for a constant energy barrier is shown in Fig. S2. In this macrospin model the most efficient critical current is obtained for $B_y$ close to (but not equal to) zero, for which the spin torque required for switching is $\tau_{ST}^0 \approx B_{an}^0 (|I_c|)/2$. Only a few mT of $B_y$ is enough to define a clear switching direction, making the switching deterministic (see Figs. 1a,b of the main text). In practical device geometries, this small, fixed, in-plane magnetic field can be easily applied by the magnetic dipole field of a nearby magnetic layer. The energy barrier against thermal reversal for the switching layer is then approximately $\Delta f_0 = At M_S (I=0) B_{an}^0 (I=0) / 2$, which we will set equal to $40\, k_B T$.

To complete the calculation of an optimum critical current for a thermally stable magnetic element we use the following relationships, applicable for the sample geometry we are considering.

$$\tau_{ST}^0 = \frac{\hbar J_S}{2eM_s t} \tag{S4}$$

$$J_S = J_e \left( \frac{J_S(d=\infty)}{J_e} \right) \left[ 1 - \text{sech}(d/\lambda_{sf}) \right] \quad \text{(explained in Section S1 above)} \tag{S5}$$

$$\frac{J_{e,Pt}}{\sigma_{Pt}} = \frac{J_{e,F}}{\sigma_F} \tag{S6}$$

$$I = J_{e,Pt} w d + J_{e,F} w t. \tag{S7}$$

Here $d$ is the thickness of the Pt layer, $w$ is the sample width (in the $\hat{x}$ direction, perpendicular to the current) and we will use that $L$ is the sample length (in the $\hat{y}$ direction, parallel to the current). We have taken into account that the electrical conductivities of the Pt ($\sigma_{Pt}$) and the ferromagnetic layer ($\sigma_F$) may be different, so the charge current densities in the two layers, $J_{e,Pt}$ and $J_{e,F}$, may also differ. Assembling these equations, the condition $\tau_{ST}^0 \approx B_{an}^0(|I_c|)/2$ is equivalent to

$$I_c = \frac{2e(40\,k_B T)\left[d + (\sigma_F/\sigma_{Pt})t\right]}{\hbar L \left(J_S(d=\infty)/J_{e,Pt}\right)\left[1 - \text{sech}(d/\lambda_{sf})\right]} \frac{M_S(|I_c|) B_{an}^0(|I_c|)}{M_S(I=0) B_{an}^0(I=0)}. \tag{S8}$$

We assume $t = 0.6$ nm, $d = 2.0$ nm, $L = 200$ nm, $J_S(d=\infty)/J_{e,Pt} = 0.07$[S5], and $\lambda_{sf} = 1.4$ nm for Pt[S5]. We also assume for simplicity that $\sigma_{Pt} = \sigma_F$. The energy barrier of $40\,k_B T$ corresponds, e.g., to a perpendicular anisotropy field $B_{an}^0(I=0) \approx 28$ mT for a sample with $M_S = 1.0 \times 10^6$ A/m, $L = 200$ nm, $w = 100$ nm. If we ignore the effects of heating, so that $M_S(|I_c|) B_{an}^0(|I_c|) / \left[M_S(I=0) B_{an}^0(I=0)\right] = 1$, then based on Eq. (S8) we estimate a critical current $I_c \sim 170$ µA with no thermal assistance. We have assumed that the energy barrier scales with the sample area, so the dependence of $I_c$ on $1/L$ in Eq. (S8) reflects a dependence on the aspect ratio of the device; $I_c$ can be reduced further by increasing $L$ beyond 200 nm while reducing $w$ to keep $A = Lw$ constant, as long as the magnetization dynamics can still be described in the macrospin approximation. $I_c$ is also likely to be decreased further if there is any heating, due to the reduction of the ratio $M_S(|I_c|)B_{an}^0(|I_c|)/[M_S(I=0)B_{an}^0(I=0)]$. Heating need not compromise the energy barrier for thermal stability, since stability is required only when $I = 0$. Even further reductions of the switching current might be achieved via advances in materials research that increase $J_S(d=\infty)/J_e$ and/or reduce the value of $\lambda_{sf}$ compared to pure Pt, or by taking advantage of non-macrospin dynamics during the switching process. We therefore conclude that spin Hall torque switching of an optimized perpendicularly magnetized memory element should be possible with currents competitive with the optimum currents required for switching driven by conventional spin transfer torque in magnetic tunnel junctions[S8-S10].

We also note that we have considered only the case of perpendicularly-magnetized magnetic films. The spin Hall torque should also be capable of producing reversal of an in-plane magnetized film. In that case the condition defining the switching current is not be that the spin Hall torque overcomes an anisotropy field, but rather that the spin Hall torque contributes an effectively negative damping that overcomes the positive Gilbert damping coefficient of the magnetic film (similar to the case of conventional spin torque tunnel junction in which both the polarizer and free layer are in-plane polarized[S7]). In this case, for an in-plane polarized film, the fact that the Pt in a Pt/ferromagnet bilayer significantly increases the Gilbert damping relative to the intrinsic damping of a thin magnetic layer[S11] will, we suspect, increase the critical currents for spin-Hall switching sufficiently to make them uncompetitive with existing technologies.

However, if materials are identified which have large spin Hall effects but which do not increase the magnetic damping of an adjacent magnetic film, SHE torque manipulation of in-plane-magnetized films may also be a promising strategy for magnetic memory devices or nonvolatile logic.

### S4. Comments on arguments by Miron et al. in ref. S1

In the Supplementary Information for their article in ref. S1, Miron et al. made two arguments why they believe that the current-induced switching they measured cannot be explained by a spin Hall torque.

The first argument concerns the slopes of the switching boundaries in data like Fig. 3d of our main text, which allow comparison between the relative strength of the current and the $\hat{z}$-component of magnetic field in giving rise to switching transitions. Using an in-plane field of 100 mT (in the $\hat{y}$ direction by our convention) and sweeping $B_z$, Miron et al. observed slopes corresponding to $|dB_z/dJ_e| \approx$ 7-9 $\times$ 10$^{-10}$ T/(A/cm$^2$). They argued that this was too strong a shift to be explained by the spin Hall torque, which they estimated to have a maximum possible value $d\tau_{ST}^0/dJ_e =$ 1.6 $\times$ 10$^{-10}$ T/(A/cm$^2$). Indeed, we calculate that in a simple zero-temperature macrospin model one should have $|dB_z/d\tau_{ST}^0| = |dB_z/dJ_e|/|d\tau_{ST}^0/dJ_e| \approx 2$ when $B_y \ll B_{an}^0$ (see Fig. 3b of the main text), so that if heating can be neglected then $|dB_z/dJ_e|$ should be no more than a factor of two greater than $d\tau_{ST}^0/dJ_e$.

We observe a similar, large difference between the measured slopes of the switching boundaries in our data and our measurement of $\tau_{ST}^0$. The slopes in Fig. 3d of the main text correspond to $|dB_z/dI| \approx$ 1.3 mT/mA or $|dB_z/dJ_e| \approx 7 \times 10^{-10}$ T/(A/cm$^2$). Like the result in ref.

S1, this is more than a factor of two greater than the value of the spin Hall torque that we measure independently, $d\tau_{ST}^0/dI = 0.33$ mT/mA or $d\tau_{ST}^0/dJ_e = 1.7 \times 10^{-10}$ T/(A/cm$^2$). Nevertheless, as we show in Fig. 3d of the main text, the switching boundaries that we measure are still in quantitative agreement with the predictions of our modified Stoner-Wohlfarth model. The reason why the modified Stoner-Wohlfarth model agrees with the data while the zero-temperature macrospin model does not is that the modified Stoner-Wohlfarth model takes into account, at least approximately, the effects of heating and spatially incoherent magnetization reversal, in addition to the spin Hall torque. The consequences of heating, to reduce $B_{an}^0$ and $B_c$ with increasing $|I|$, make the current increasingly effective in contributing to switching as $|I|$ is increased, and therefore shift the slopes of the switching boundaries $|dB_z/dJ_e|$ to larger values.

We should note, to be clear, that heating alone cannot explain the measured current-induced switching phenomena in the absence of a spin Hall torque, since heating alone cannot explain the strong dependence of the switching direction on the sign of the current seen in Fig. 1 of the main text.

The second argument of Miron et al. against the spin Hall torque mechanism was based on measurements of three sets of Pt/Co/AlO$_x$ layers formed by oxidizing Al layers with different thicknesses. Samples made with thicker Al layers, so that they were less oxidized, exhibited lower values of $B_{an}^0(I=0)$ and $B_c(I=0)$, but higher critical currents for switching. Miron et al. argued that if the spin Hall torque were the mechanism for switching and if the strength of the spin Hall torque were the same in the different samples, then the samples with the lower values of $B_{an}^0(I=0)$ and $B_c(I=0)$ must have lower critical currents, in conflict with the data. We find this argument unpersuasive. First, the different samples are unlikely to have the same strength of the spin Hall torque. The thicker, less-oxidized samples will have either partially unoxidized metallic aluminum or less oxidation of the Co layer (or both) compared to the thinner, more oxidized samples, with the result that in the less-oxidized samples a smaller share of the applied current will flow through the Pt layer, due to shunting through the Co or Al. This will decrease the current density in the Pt and hence the strength of the spin Hall torque in the thicker samples and therefore increase the total critical current needed for switching. Second, the amount of heating in these samples is substantial. The thicker, less-oxidized samples will have lower resistances and therefore somewhat less heating. Since $B_{an}^0$ and $B_c$ are temperature dependent, this means that $B_{an}^0(I = I_c)$ and/or $B_c(I = I_c)$ may be greater in the less-oxidized samples than in the more-oxidized samples even if the reverse is true for $I = 0$. This factor could also contribute to larger critical currents for the less-oxidized samples. Finally, different extents of formation of antiferromagnetic Co oxide at the Co/Al interface and the fluctuating exchange biasing[S12] might affect how readily a current promotes the nucleation and motion of magnetic domain walls during the magnetic reversal process and therefore alter the values of critical currents.

**S5. Determining the zero point for the applied magnetic field angle $\beta$.**

Using a rotary sample stage that provides a 360° rotation with 0.02° precision, we recorded magnetization curves ($R_H$ vs. $B_{ext}$) for different fixed values of the field angle $\beta$. Small currents ($I$ = 1 mA) were utilized for those measurements in order to minimize the effect of spin torque on the magnetization. The curves shown in Figures S3a and S3b were obtained near $\beta \approx 0$ with a difference in $\beta$ of less than 1°. It can be seen that the sign of $R_H$ is opposite between the $R_H$ vs. $B_{ext}$ curves of these two plots, indicating that the z components of $B_{ext}$ for those two angles are positive and negative, separately. We are therefore able to determine that the angle corresponding to $\beta = 0$ is bounded by the two positions.

**S6. Determining the phase diagram in the macrospin model for a spin Hall torque acting on a magnetic layer with perpendicular magnetic anisotropy.**

We will consider the case, for simplicity, that there is no applied magnetic field component in the $x$ direction (as defined in Fig. 1e of the main text), and we will determine the magnetic orientations $\hat{m}(\tau_{ST}^0, B_y, B_z)$ that satisfy the torque balance equation $\vec{\tau}_{tot} \equiv \vec{\tau}_{ST} + \vec{\tau}_{ext} + \vec{\tau}_{an} = 0$, where

$$\vec{\tau}_{ST} = \tau_{ST}^0 (\hat{m} \times \hat{x} \times \hat{m}) \tag{S9}$$

$$\vec{\tau}_{ext} = -\hat{m} \times \vec{B}_{ext} \tag{S10}$$

$$\vec{\tau}_{an} = -\hat{m} \times \vec{B}_{an} = -\hat{m} \times \left[ -B_{an}^0 (\hat{m} - m_z \hat{z}) \right] = -B_{an}^0 m_z (\hat{m} \times \hat{z}). \tag{S11}$$

Defining the $x, y, z$ axes as in Fig. 1e of the main text, we evaluate each of these torques in Cartesian coordinates, using $\hat{m} = (m_x, m_y, m_z)$ with $m_x^2 + m_y^2 + m_z^2 = 1$ and $\vec{B}_{ext} = (B_x, B_y, B_z)$.

$$\vec{\tau}_{tot} = 0 = \begin{aligned} &\hat{x}\left[ \tau_{ST}^0 (m_y^2 + m_z^2) - B_z m_y + B_y m_z - B_{an}^0 m_y m_z \right] \\ &+ \hat{y}\left[ -\tau_{ST}^0 m_x m_y + B_z m_x - B_x m_z + B_{an}^0 m_x m_z \right] \\ &+ \hat{z}\left[ -\tau_{ST}^0 m_x m_z - B_y m_x + B_x m_y \right] \end{aligned} \tag{S12}$$

By construction, all of the torques are perpendicular to $\hat{m}$, so it is convenient to consider just the two components within the plane perpendicular to $\hat{m}$ by taking projections along the directions

$$\hat{x} \times \hat{m} = -m_z \hat{y} + m_y \hat{z} \tag{S13}$$

$$\text{and} \quad \hat{m} \times \hat{x} \times \hat{m} = (m_y^2 + m_z^2)\hat{x} - m_x m_y \hat{y} - m_x m_z \hat{z}. \tag{S14}$$

It is reasonable to perform these projections for all cases except when $\hat{m} = \hat{x}$ (in which case both vectors are zero). This case can be ignored because $\hat{m} = \hat{x}$ is not a stable solution of Eq. (S12) for any interesting physical case with $B_x = 0$. These two projections give the results

$$(\hat{x} \times \hat{m}) \cdot \vec{\tau}_{tot} = 0 = B_x \left(m_y^2 + m_z^2\right) - B_y m_x m_y - B_z m_x m_z - B_{an}^0 m_x m_z^2 \tag{S15}$$

$$(\hat{m} \times \hat{x} \times \hat{m}) \cdot \vec{\tau}_{tot} = 0 = \tau_{ST}^0 \left(m_y^2 + m_z^2\right) + B_y m_z - B_z m_y - B_{an}^0 m_y m_z. \tag{S16}$$

If we consider only cases in which $B_x = 0$, we have

$$0 = m_x \left(B_y m_y + B_z m_z + B_{an}^0 m_z^2\right) \tag{S17}$$

$$0 = \tau_{ST}^0 \left(m_y^2 + m_z^2\right) + B_y m_z - B_z m_y - B_{an}^0 m_y m_z. \tag{S18}$$

Given the form of Eq. (S17), the solutions for the case $B_x = 0$ break up into two classes, either $m_x = 0$ or $B_y m_y + B_z m_z + B_{an}^0 m_z^2 = 0$, corresponding to different phases in which the magnetization remains in the $yz$ plane (at small values of $\tau_{ST}^0$) or in which it is forced to tilt out of this plane (at larger values of $\tau_{ST}^0$). We will consider these two phases separately.

***Solutions for $B_x = 0$ with $m_x = 0$***: In this case, we have $m_y^2 + m_z^2 = 1$ and Eq. (S18) becomes

$$0 = \tau_{ST}^0 + B_y m_z - B_z m_y - B_{an}^0 m_y m_z. \tag{S19}$$

This is identical to Eq. (1) in the main text with the identifications that for $m_x = 0$ then $m_y = \cos\theta$, $m_z = \sin\theta$, $B_y = B_{ext} \cos\beta$, and $B_z = B_{ext} \sin\beta$. Equation (S19) can be solved numerically. The stable and unstable equilibrium states can be distinguished by simulating the LLG equation with positive damping coefficient.

For convenience in determining the limits of stability for the $m_x = 0$ states (analogous to the Stoner-Wohlfarth astroid), we can rewrite Eq. (S19) in the form

$$0 = \tau_{tot}^x \equiv \tau_{ST}^0 + B_y \sin\theta - B_z \cos\theta - B_{an}^0 \sin\theta \cos\theta. \tag{S20}$$

Within a zero-temperature macrospin model, the switching conditions for which an $m_x = 0$ solution becomes unstable are determined by Eq. (S20) together with the condition

$$d\tau_{tot}^x / d\theta = 0 = B_y \cos\theta + B_z \sin\theta + B_{an}^0 \left(\sin^2\theta - \cos^2\theta\right). \tag{S21}$$

We can reach a symmetric form for the switching conditions by combining Eqs. (S20) and (S21) to eliminate first $B_z$ and then $B_y$. To eliminate $B_z$, we multiply Eq. (S20) by $\sin\theta$ and Eq. (S21) by $\cos\theta$ and then add, to give

$$\tau_{ST}^0 \sin\theta + B_y - B_{an}^0 \cos^3\theta = 0. \tag{S22}$$

To eliminate $B_y$, we multiply Eq. (S20) by $-\cos\theta$ and Eq. (S21) by $\sin\theta$ and then add, to give

$$-\tau_{ST}^0 \cos\theta + B_z + B_{an}^0 \sin^3\theta = 0. \tag{S23}$$

Equations (S22) and (S23) are identical to Equations (2) and (3) in the main text. We determined the switching boundaries plotted in Figures 3a and 3b of the paper by solving Eqs. (S22) and (S23) numerically.

**Solutions for** $B_x = 0$ **with** $m_x \neq 0$ **but with** $B_y m_y + B_z m_z + B_{an}^0 m_z^2 = 0$: From Eq. (S17) we have

$$m_y = -\frac{B_z m_z + B_{an}^0 m_z^2}{B_y}. \tag{S24}$$

Substituting this into Eq. (S18) and factoring the result, we find

$$0 = m_z \left(\tau_{ST}^0 m_z + B_y\right) \left[\frac{\left(B_{an}^0\right)^2}{B_y^2} m_z^2 + \frac{2 B_z B_{an}^0}{B_y^2} m_z + \frac{B_z^2}{B_y^2} + 1\right]. \tag{S25}$$

There is no allowed physical solution for which the quadratic equation in the square brackets is equal to zero (the discriminant is negative). There is an apparent solution with $m_z = 0$, { $m_x = 1$, $m_y = 0$, $m_z = 0$ }, but this arises as an artifact of my choice to project along the axes $\hat{x} \times \hat{m}$ and $\hat{m} \times \hat{x} \times \hat{m}$ -- these vectors are both trivially zero if $\hat{m} = \hat{x}$. By substituting directly into the starting equation, Eq. (S12), one can see that $\{m_x = 1, m_y = 0, m_z = 0\}$ is not actually a solution of the starting equation. Therefore the only allowed solution in this class is

$$m_z = -\frac{B_y}{\tau_{ST}^0} \tag{S26}$$

$$m_y = -\frac{B_{an}^0 B_y}{(\tau_{ST}^0)^2} + \frac{B_z}{\tau_{ST}^0} \tag{S27}$$

$$m_x = \pm\sqrt{1-(m_y)^2-(m_z)^2} = \pm\sqrt{1-\left(\frac{B_y}{\tau_{ST}^0}\right)^2 - \left(-\frac{B_{an}^0 B_y}{(\tau_{ST}^0)^2} + \frac{B_z}{\tau_{ST}^0}\right)^2}. \tag{S28}$$

Only one sign of $m_x$ will be a stable equilibrium. We determine which one is stable by simulating the equation of motion $(1/|\gamma|)d\hat{m}/dt = \vec{\tau}_{tot} + (\alpha/|\gamma|)\hat{m}\times(d\hat{m}/dt)$ with $\alpha > 0$. The limits of stability for these high-current $m_x \neq 0$ solution (the dotted lines in Figs. 3a,b in the main text) are determined by the condition that the quantity under the square root in Eq. (S28) changes from positive to negative.

We have verified these two different classes of analytical solutions ($m_x = 0$ and $m_x \neq 0$) as well by numerical solution of the macrospin Landau-Lifshitz-Gilbert equation with the spin Hall torque included. There are no dynamical states that can be excited by a constant spin Hall torque within this model for a magnetic film with perpendicular anisotropy and for $B_x = 0$.

The full phase diagram that we calculate for $\hat{m}(\tau_{ST}^0, B_y, B_z)$ is available in the accompanying text file named 'plot3dphase.m'. This is a Matlab program file which can be executed and plotted within Matlab to visualize different aspects of the phase diagram. The *x*, *y* and *z* axes of the figure correspond to $B_y$, $\tau_{ST}^0$ and $B_z$. The horizontal and vertical transparent planes depicted in the 3-D image correspond to the sections drawn in Fig. 3a and Fig. 3b in the main text.

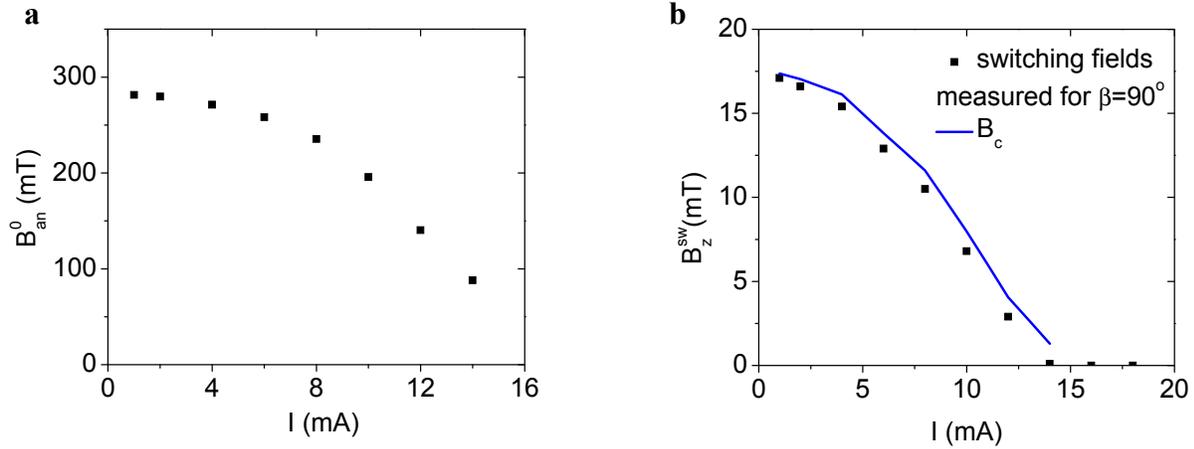

**Figure S1. The dependence of the perpendicular anisotropy field and the intrinsic coercive field on the applied current. a,** $B_{an}^0$ as a function of the applied current $|I|$ measured as explained in Section S2. **b,** Square points: Measured switching fields $B_z^{sw}$ for field sweeps perpendicular to the sample plane. Solid line: The intrinsic coercive field $B_c(|I|)$ after correcting for the effects of the spin Hall torque as explained in Section S2.

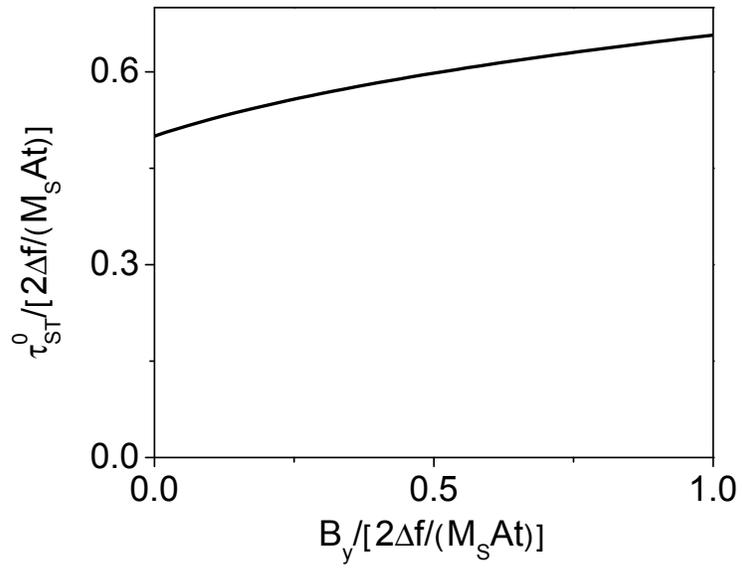

**Figure S2. The dependence of spin torque required for switching on $B_y$, for a constant energy barrier against thermal fluctuations.** In this calculation, $B_{an}^0$ is adjusted as a function of $B_y$ to keep the energy barrier constant, as discussed in Section S3.

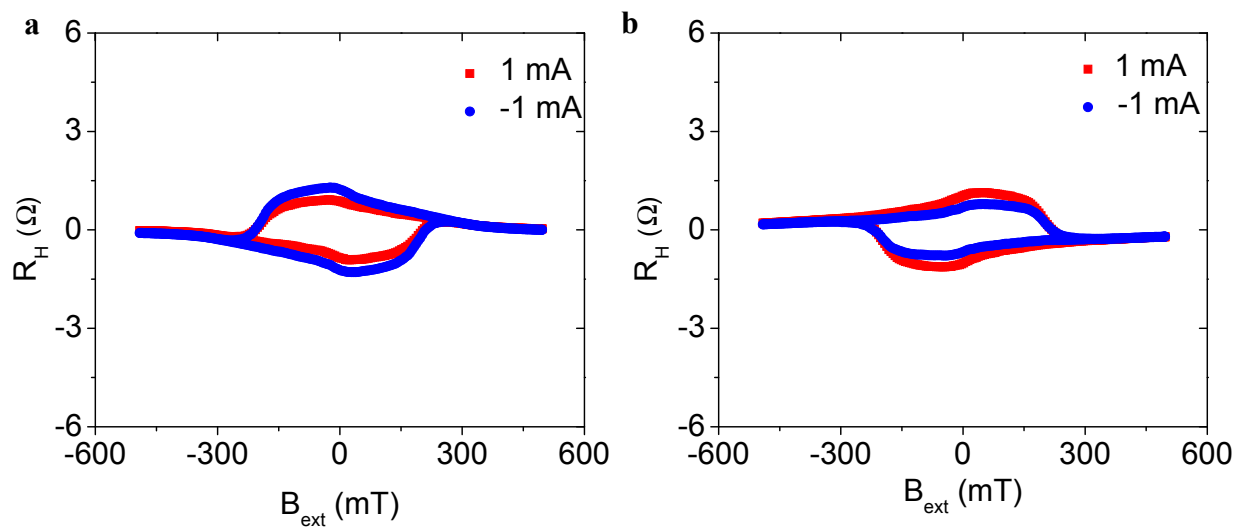

**Figure S3. Hysteresis loops used to determine the zero point for the angle β.** The curves shown in **a** and **b** correspond to a small positive and negative *β*, respectively.

**Supplementary Information Section References**: